\documentclass{emulateapj}
\usepackage{amsmath}
\usepackage{amssymb}
\usepackage{amsthm}
\usepackage{array}
\usepackage{float}
\usepackage{graphicx}
\usepackage{subfigure}

\newcommand{\myemail}{kzk15@psu.edu}

\slugcomment{}

\shorttitle{Galaxy Mergers as a Source of CRs, $\nu$s, and $\gamma$s}
\shortauthors{Kashiyama \& M$\rm \acute{e}$sz${\rm \acute{a}}$ros}

\begin{document}

\title{Galaxy Mergers as a Source of Cosmic Rays, Neutrinos, and Gamma Rays}

\author{Kazumi Kashiyama and Peter M$\rm \acute{e}$sz${\rm \acute{a}}$ros}
\affil{Center for Particle \& Gravitational Astrophysics; Department of Physics; Department of Astronomy \& Astrophysics; \\ Pennsylvania State University, University Park, PA 16802, USA}
\email{\myemail}

\begin{abstract}
We investigate the shock acceleration of particles in massive galaxy mergers or collisions, 
and show that cosmic rays~(CRs) can be accelerated up to the second knee energy $\sim 0.1\mbox{-}1$ EeV and possibly beyond, with a hard spectral index of $\Gamma \approx 2$. 
Such CRs lose their energy via hadronuclear interactions within a dynamical timescale of the merger shock, 
producing gamma rays and neutrinos as a by-product.  
If $\sim 10 \%$ of the shock dissipated energy goes into CR acceleration, 
some local merging galaxies will produce gamma-ray counterparts detectable by CTA. 
Also, based on the concordance cosmology, where a good fraction of the massive galaxies experience a major merger in a cosmological timescale,
the neutrino counterparts can constitute $\sim 20\mbox{-}60 \ \%$ of the isotropic background detected by IceCube. 
\end{abstract}

\keywords{acceleration of particles --- neutrinos --- shock waves --- galaxies: evolution}

\section{Introduction}
The origin of cosmic rays~(CRs), in particular above the knee energy $\gtrsim 10^{16} \ \rm eV$
and at ultra-high energies~(UHEs; $\gtrsim 10^{19} \ \rm eV$), is still uncertain, 
but the discovery of sub-PeV neutrinos by IceCube~\citep{Aartsen_et_al_2013, IceCube_2013,IceCube_2014} may provide clues to this origin.
Given that the distribution of the arrival directions is consistent with isotropy,
most events are likely to come from extragalactic PeV accelerators~\citep[][and references therein]{Murase_et_al_2013}, 
although a fraction of them may be attributed to Galactic sources~\citep[e.g.,][]{Fox_et_al_2013, Razzaque_2013, Ahlers_Murase_2013}. 

Relevant constraints on the parent CR acceleration and the isotropic-neutrino-background~(INB) production were given by \cite{Murase_et_al_2013}.
They showed that if $pp$ interactions are responsible for the INB, the parent CR spectrum must have a hard spectral 
index $\Gamma \lesssim 2.2$ in order for the by-product gamma rays not to overshoot the 
observed isotropic gamma-ray background~(IGB)~\citep{Fermi_2010}. 
Conversely, a significant fraction of the INB and the IGB may be attributed 
to a single type of source with a sufficiently hard spectral index. 
For this, the required local CR injection rate below the knee energies per energy decade is 
$\varepsilon_{cr} Q_{\varepsilon_{cr}} \sim 10^{44} \ \min [1, f_{pp}]^{-1} \ \rm erg \ Mpc^{-3}  
\ yr^{-1}$, where $f_{pp}$ is the effective hadronuclear optical depth at the source. 

Importantly, the above injection rate is comparable to that required for the UHECR sources 
$\varepsilon_{cr} Q_{\varepsilon_{cr}} \sim 0.5 \times 10^{44} \ \rm erg \ Mpc^{-3}  \ yr^{-1}$,
or equivalently, the observed flux of the INB is consistent with the 
Waxman-Bahcall limit~\citep{Waxman_Bahcall_1999} with $f_{pp} \sim 1$. 
This may indicate that UHECRs also originate in the same type of sources~\citep[e.g.,][]{Katz_et_al_2013}. 
Note, however, that if indeed the parent CR spectrum of the INB is as hard as 
$\Gamma \approx 2$ and the maximum energy is in the UHE range,  $f_{pp}$ may have to become low 
at $\gtrsim 10^{17} \ \rm eV$ in order for the by-product neutrinos to be consistent with the 
non-detection of super-PeV neutrinos by IceCube~\citep{Laha_et_al_2013}.

In summary, a significant fraction of the INB, the IGB, and possibly also the UHECRs, 
can be consistently attributed to a single type of source based on a $pp$ scenario if 
\begin{itemize}
\item[(i)] the sources have the right local CR injection rate of $Q_{cr} \equiv \int d 
\varepsilon_{cr} Q_{\varepsilon_{cr}} \gtrsim 10^{45} \ \rm erg \ Mpc^{-3}  \ yr^{-1}$, 
\item[(ii)] the sources accelerate CRs up to UHEs with a hard spectral index of $\Gamma \lesssim 2.2$, 
and
\item[(iii)] the CRs with $\lesssim 100 \ \rm PeV$ lose a fraction of their energy via hadronuclear reactions, i.e., $f_{pp} \sim 1 \ Q_{cr, 45}$,  
\end{itemize}
(hereafter, we use $Q_x = Q/10^x$). 
Individual local source identifications using e.g., HAWC~\citep{HAWC_2012} and CTA~\citep{CTA_2011} are the keys to testing the scenarios.    

Here, we investigate galaxy merger shocks~(GMSs) as such a possibility. 
In the concordance cosmology, a good fraction of the massive galaxies experience a major merger 
and several minor mergers in a cosmological timescale. 
During the direct encounters between two galaxies, their cold-gas components  are shocked 
with sufficiently large Mach numbers to enable an efficient CR acceleration. 
Below, we show that GMSs can meet the conditions (ii) and (iii), and the anticipated CR injection 
rate can be a good fraction of that required by condition (i) for a CR acceleration efficiency 
of $\sim 10 \ \%$\footnote{We should note that the GMSs have been investigated as the source of 
UHECRs~\citep[e.g.,][]{Cesarsky_Ptuskin_1993, Jones_1998} and gamma rays~\citep{Lisenfeld_Volk_2010}, 
separately.}.

\section{Galaxy Merger Shock Scenario}
\subsection{Energy Budget}
First, let us estimate the intrinsic energy budget of a galaxy merger. 
Hereafter, we mainly consider major mergers between massive galaxies with a stellar mass of $M_\ast \sim 10^{11} M_\odot$, 
which likely dominate the total energy budget.  
The energy dissipated by a GMS can be estimated as $\overline{E}_{gms} \approx 0.5 M_{gas} v_s^2$, or
\begin{equation}
\overline{E}_{gms} \sim 2.5 \times 10^{58} \  M_{gas, 10} v_{s, 7.7}{}^2 \ \rm erg. 
\end{equation}
where $M_{gas}$ is the shocked gas mass and $v_s$ is the shock velocity.  
At least up to $z \sim 3$, the galaxy gas mass is typically $\sim 10 \%$ of the stellar mass 
at the high-mass end with $M_\ast \gtrsim 10^{11} \ \rm M_\odot$~\citep[e.g.][]{Conselice_et_al_2013}. 
The shock velocity is essentially the relative velocity between the merging galaxies, 
which can range up from $v_s \sim (3\mbox{-}9) \times 10^7 \ \rm cm \ s^{-1}$ at a pericenter distance of $R_{gal} \lesssim 1\mbox{-}10 \ \rm kpc \sim 3.1 \times 10^{(21\mbox{-}22)} \ \rm cm$ 
in the point particle approximation.
The shock dissipation occurs with a dynamical timescale, 
\begin{equation}\label{eq:t_dyn}
t_{dyn} \approx R_{gal}/v_s \sim 20 \ R_{gal, 22.5} v_{s, 7.7}{}^{-1} \ \rm Myr, 
\end{equation} 
and the energy dissipation rate per merger is 
\begin{equation}
L_{gms} \sim 5.0 \times 10^{43}  \  \overline{E}_{gms, 58.5} v_{s, 7.7} R_{gal, 22.5}{}^{-1} \ \rm erg \ s^{-1}.
\end{equation}

The local major merger rate of massive galaxies has been estimated as 
${\cal R}_{gms} \gtrsim 10^{-4} \ \rm \ Mpc^{-3} \ Gyr^{-1}$~\citep[e.g.,][]{Hopkins_et_al_2010,Lotz_et_al_2011}. 
Equivalently, an $\gtrsim 10 \ \%$ of massive galaxies experience major mergers in 
a cosmological time given that the mean density is $n_{gal} \sim 10^{-2} \ \rm Mpc^{-3}$~\citep{Bell_et_al_2003}.
The possible CR injection rate by GMSs can be estimated as $Q_{cr, gms} \approx \xi_{cr} \overline{E}_{gms} {\cal R}_{gms}$, or 
\begin{eqnarray}\label{eq:inj_rate}
Q_{cr, gms} \sim 3.2 &\times& 10^{44} \ {\rm erg \ Mpc^{-3}  \ yr^{-1}} \notag \\
&\times& \xi_{cr,-1} \overline{E}_{gms, 58.5} {\cal R}_{gms, -4}, 
\end{eqnarray}
where $\xi_{cr}$ is the CR acceleration efficiency.

\subsection{Cosmic-Ray Acceleration}
\begin{figure}
\includegraphics[width=90mm]{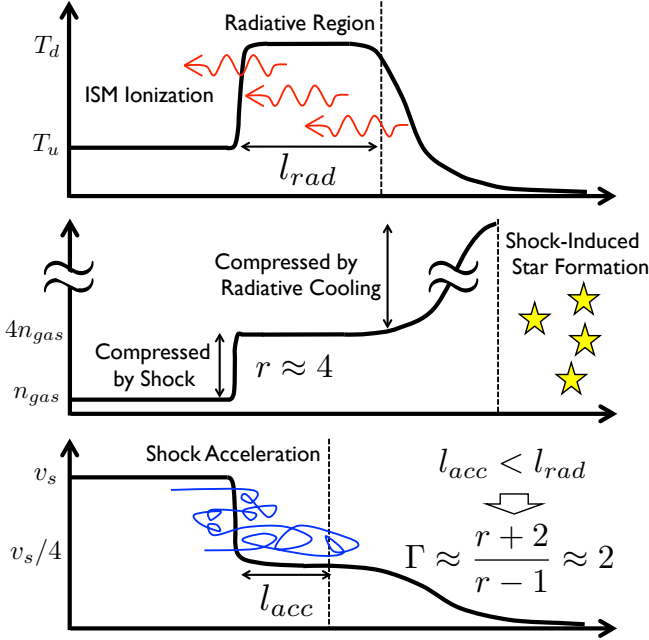}
\caption{Schematic picture of GMS and the DSA in situ; temperature (top), density (middle), and velocity in the shock rest frame (bottom).}
\label{fig1}
\end{figure}
Next, we consider the diffusive shock acceleration (DSA) mechanism~\citep[e.g.,][]{Drury_1983} at GMSs. 
For this, we consider first the characteristics of GMS (see Fig. \ref{fig1}). 
The Mach number of GMSs can be estimated as ${\cal M} \approx v_s/(5k_B T_u/3\mu)^{1/2} \sim 33 \ v_{s, 7.7} T_{u, 4}{}^{1/2}$, 
which is much larger than unity for the cold gas component with $T_u < 10^4 \ \rm K$. 
Here, we use $\mu \approx 0.61m_p$ for ionized gas (see below). 
On the other hand, the temperature downstream of the shock can be estimated as $T_d \approx (3/16)(\mu v_s^2/k_B) \sim 3.4 \times 10^6 \ v_{s, 7.7}{}^2 \ \rm K$. 
Such a high downstream temperature plays an important role in characterizing GMSs, 
first of all because the radiative cooling becomes relevant in this region.  
At $T_d \gtrsim 10^6 \ \rm K$, the main cooling channel is provided by metal line emissions and free-free emission~\citep[e.g.,][]{Sutherland_Dopita_1993}. 
Using an approximate form of the cooling function for a solar metallicity interstellar medium~(ISM), the cooling timescale can be estimated as 
\citep{Drain_2011} 
\begin{equation}\label{eq:t_rad}
t_{rad} \sim  1.6 \  n_{gas, 0}{}^{-1} v_{s, 7.7}{}^{17/5} \ \rm Myr, 
\end{equation}
where $n_{gas}$ is the preshock gas density. 
From Eqs. (\ref{eq:t_dyn}) and (\ref{eq:t_rad}), one can see that for a shock velocity $v_s \lesssim 8.8 \times 10^7 \ R_{gal, 22.5}{}^{5/22} n_{gas, 0}{}^{5/22} \ \rm cm \ s^{-1}$, 
the downstream plasma cools within a dynamical timescale, i.e., the GMS shocks are radiative.  
In this case, a larger compression ratio than the $r=4$ ratio of strong adiabatic shocks is realized beyond the radiative region, which is 
\begin{equation}\label{eq:l_rad}
l_{rad} \approx v_s/4 \times  t_{rad} \sim 210 \ n_{gas, 0}{}^{-1} v_{s, 7.7}{}^{22/5} \ \rm pc 
\end{equation}
from the shock surface.  
Such highly compressed regions are plausible sites for violent star formation during the galaxy merger~\citep{Larson_Tinsley_1978,Barnes_2004,Saitoh_et_al_2009}.
Second, the UV photons from the radiative zone can significantly ionize the upstream ISM. 
\cite{Shull_McKee_1979} showed that this occurs for shock velocities $v_s \gtrsim 110 \ \rm km \ s^{-1}$. 
In this case, one can naturally expect Alfv$\rm \acute{e}$n-wave turbulence behind the GMS, 
which is crucial for the DSA mechanism.  
Large-scale magnetic fields in interacting galaxies are observed to be $B \sim 10 \ \rm \mu G$~\citep{Drzazga_et_al_2011}.  
One can also expect an amplification of the magnetic fields at least locally around the shocks as $B \approx (4 \pi \xi_B n_{gas} m_p v_s{}^2)^{1/2}$, or
\begin{equation}\label{eq:B}
B \sim 230 \ \xi_B^{1/2} n_{gas, 0}{}^{1/2} v_{s, 7.7} \ \rm \mu G, 
\end{equation}
where $\xi_B \leq 1$ is the amplification factor.  

In the DSA mechanism, 
the resultant CR spectral index, $\Gamma$, where $Q_{\varepsilon_{cr}} \propto \varepsilon_{cr}{}^{1-\Gamma}$, 
is characterized by the compression ratio as $\Gamma \approx (r+2)/(r-1)$ in the test-particle approximation.  
As discussed above, the compression ratio can be much larger than unity, $r \gg 1$, comparing the upstream and the far downstream beyond the radiative zone.  
On the other hand, the size of the acceleration region in the downstream can be estimated as  
\begin{equation}\label{eq:l_acc}
l_{acc} \approx D/v_s \sim 94 \ Z^{-1} \xi_B^{-1/2} n_{gas, 0}{}^{-1/2} v_{s, 7.7}{}^{-2} \varepsilon_{cr, 17} \ \rm pc.
\end{equation}
Here, $D \approx c R_L/3$ is the diffusion coefficient in the Bohm limit, 
$R_L = \varepsilon_{cr}/ZeB \sim 0.47 \ Z^{-1} \xi_B^{-1/2} n_{gas, 0}{}^{-1/2} v_{s, 7.7}{}^{-1} \varepsilon_{cr, 17} \ \rm pc$ is the Larmor radius, and $Z$ is the electric charge of the CRs. 
From Eqs. (\ref{eq:l_rad}) and (\ref{eq:l_acc}), the CRs in the GMS acceleration reside in the region where $r \approx 4$ 
for $v_s \gtrsim 4.4 \times 10^7 \ \xi_B{}^{-5/64} n_{gas, 0}{}^{5/64} (\varepsilon_{cr, 17}/Z)^{5/32} \ \rm cm \ s^{-1}$, 
and thus, the CR spectral index can be $\Gamma \approx 2$. 

The maximum energy possible for CRs is given by $t_{acc} = t_{dyn}$, where $t_{acc} \approx \eta D/v_s^2$ is the acceleration timescale. 
This gives $\varepsilon_{cr, max} \approx \eta Z e B R_{gal} v_s/c$, or 
\begin{equation}\label{eq:e_max}
\varepsilon_{cr, max} \sim  3.5 \times 10^{18} \ \eta Z \xi_B{}^{1/2} n_{gas, 0}{}^{1/2} v_{s, 7.7}{}^2  R_{gal,22.5} \ \rm eV.
\end{equation}
Here, $\eta$ depends on the effective compression ratio and the magnetic field configuration, etc.  
For an $r = 4$ parallel shock in the Bohm limit, $\eta = 3/20$.   
Thus, one can expect $\varepsilon_{p, max} \gtrsim 10^{17} \ \rm eV$ for proton CRs ($Z = 1$) at GMSs.    

If highly efficient magnetic-field amplification, $\xi_B \sim 1$, is the case at GMSs,
the maximum energy could be further boosted up to UHEs for, e.g., iron CRs ($Z = 26$), 
which may be consistent with the possible transition in the Auger UHECR composition reported by \cite{Auger_2010} (but see also \cite{HiRes_2010}).  
Note, however, that such UHECR fluxes can be suppressed by dissociation of the nucleus during the acceleration~\citep{Murase_Beacom_2010}. 
Also, even proton CRs can be accelerated up to UHEs with larger GMS velocities as possible in galaxy collisions, $v_s \gtrsim 10^{8} \ \rm cm \ s^{-1}$, where the shocks are adiabatic.    
However, the size of the acceleration region of proton CRs with $\varepsilon_p \gtrsim 10^{17} \ \rm eV$ can be larger than the width of the merging galaxy $\sim 0.1\mbox{-}1 \ \rm kpc$ (see Eq. \ref{eq:l_acc}). 
Then, a fraction of such CRs escape from the acceleration region, which also suppresses the UHECR flux. 

\subsection{Neutrinos and Gamma Rays}
Next, let us discuss the energy loss of the CRs, which we have so far neglected, focusing on proton CRs in this section.  
When a GMS completes sweeping the galaxies, the high-pressure radiative region begins to cool down almost adiabatically, and so do the CRs trapped in this region.  
This occurs on a dynamical timescale, $t_{dyn}$ (Eq.(\ref{eq:t_dyn})). 
On the other hand, the energy loss timescale due to $pp$ interactions can be estimated as 
\begin{equation}\label{eq:t_pp}
t_{pp} \approx 1/\kappa_{pp} n_{gas} \sigma_{pp} c  \sim  26 \ n_{gas, 0}{}^{-1} \ \rm Myr, 
\end{equation}
where $\sigma_{pp} \sim 8 \times 10^{-26} \ \rm cm^2$ and $\kappa_{pp} \sim 0.5$ for $\varepsilon_{cr} = 10^{17} \ \rm eV$, 
and the cross section evolves only logarithmically with CR energy. 
The effective hadronuclear optical depth can be estimated as $f_{pp} = t_{dyn}/t_{pp}$, or 
\begin{equation}
f_{pp} \sim 0.77 \  R_{gal, 22.5} v_{s, 7.7}{}^{-1} n_{gas, 0}.
\end{equation}
Thus, CRs predominately lose their energy via $pp$ interactions.  
Note that other energy loss processes, such as synchrotron, are irrelevant at least up to $\varepsilon_{p} \sim 10^{19} \ \rm eV$.   

In $pp$ interactions, the charged and neutral pions are produced with a ratio of $\pi_+ : \pi_{0} \approx 2 : 1$. 
The charged pions finally decay into three neutrinos and one positron with roughly the same energy, and the neutral pions decay into two gamma rays. 
The flavor is totally mixed via neutrino oscillation for extragalactic sources. 
The spectral index of both the neutrinos and gamma rays is approximately the same as that of the parent protons, i.e., $\Gamma \approx 2.0$ in our case. 
The high energy cutoff is $\varepsilon_{\nu_i, max} \sim 0.05 \times \varepsilon_{p, max}/(1+z) \sim 4 \  (1+z)^{-1} \varepsilon_{p, max, 17} \ \rm PeV$ for neutrinos, 
and twice larger for gamma rays, although gamma rays above $\gtrsim 10 \ \rm TeV$ from $\gtrsim 100 \ \rm Mpc$ are attenuated by the extragalactic background light. 

First, let us estimate the neutrino flux per flavor from a local GMS, that is, $\varepsilon_{\nu_i}^2 \phi_{\nu_i} \approx (1/6) \min[1, f_{pp}] (\xi_{cr}L_{gms}/{\cal C}) (1/4 \pi d_L{}^2)$, or 
\begin{eqnarray}
\varepsilon_{\nu_i}^2 \phi_{\nu_i} \sim &1.5& \times 10^{-13} \ {\rm erg \ cm^{-2} \ s^{-1} } (18/{\cal C}) (d_L/50 \ {\rm Mpc})^{-2} \notag \\ 
&\times&  f_{pp} \xi_{cr, -1} \overline{E}_{gms, 58.5} v_{s, 7.7} R_{gal, 22.5}{}^{-1}.
\end{eqnarray}
Here, $d_L$ is the luminosity distance to the source, and 
${\cal C} = (1-(\varepsilon_{p, max}/\varepsilon_{p, min})^{2-\Gamma})/(\Gamma-2)$ (or ${\cal C} = \log(\varepsilon_{p, max}/\varepsilon_{p, min})$ for $\Gamma = 2$) is the bolometric collection factor. 
We take $\varepsilon_{p, max} = 10^{17} \ \rm eV$ and $\varepsilon_{p, min} = \rm GeV$ for the estimate. 
With the above parameter set, one can only expect a small muon event rate, $\sim 0.01 \ \rm yr^{-1}$, using IceCube. 
Thus, individual local GMS source identifications through high-energy neutrinos will be difficult in the coming years.   

On the other hand, the gamma-ray flux from a local GMS can be estimated as 
$\varepsilon_{\gamma}^2 \phi_\gamma \approx 2 \times \varepsilon_{\nu}^2 \phi_\nu |_{\varepsilon_\nu = 0.5 \varepsilon_\gamma}$, or
\begin{eqnarray}
\varepsilon_{\gamma}^2 \phi_\gamma \sim &3.0& \times 10^{-13} \ {\rm erg \ cm^{-2} \ s^{-1} } (18/{\cal C}) (d_L/50 \ {\rm Mpc})^{-2} \notag \\ 
&\times&  f_{pp} \xi_{cr, -1}  \overline{E}_{gms, 58.5} v_{s, 7.7} R_{gal, 22.5}{}^{-1}. 
\end{eqnarray}
Detecting such a flux is marginally difficult for {\it Fermi} and current air Cherenkov telescopes, 
but can be detectable by CTA up to $d_L \lesssim 100 \ \rm Mpc$ with a sufficiently long observation time.  
Candidate sources are discussed in the next section. 

\begin{figure}
\includegraphics[width=90mm]{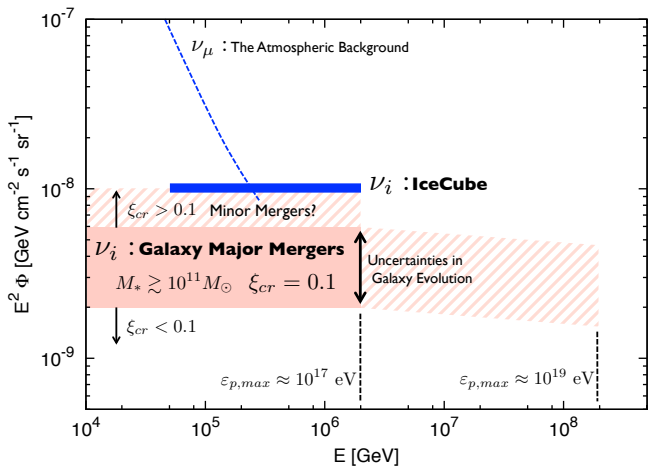}
\caption{INB flux from massive major GMSs, indicated by the shaded region. 
The striped regions show the possible extensions due to GMSs in minor mergers, or due to a more effective CR acceleration
where $\varepsilon_{p, max} = 10^{19} \ \rm eV$ corresponds to a case with $\xi_B \sim 1$ and $v_s \sim 10^8 \ \rm cm \ s^{-1}$ (see Eq. \ref{eq:e_max}). 
The solid and dashed lines represent the INB and the atmospheric background flux observed by IceCube, respectively.}
\label{fig2}
\end{figure}

Finally, let us estimate the diffuse flux from cosmological GMSs. 
Generally, the INB flux can be estimated as~\citep[e.g.,][]{Waxman_Bahcall_1999}
\begin{equation}\label{eq:nu_diffuse_flux}
\varepsilon_{\nu_i}^2 \Phi_{\nu_i} \approx \frac{c t_{H} \xi_z}{4 \pi} \frac{1}{6} \min[1,f_{pp}](\varepsilon_{cr} Q_{\varepsilon_{cr}}).  
\end{equation}
where $t_{H} \sim 13.2 \ \rm Gyr$ is the Hubble timescale with cosmological parameters $h = 0.71$, $\Omega_m = 0.3$, and $\Omega_\Lambda = 0.7$, 
$\xi_z$ accounts for cosmological evolution of the source, and $\varepsilon_{cr} Q_{\varepsilon_{cr}} = Q_{cr}/{\cal C}$.
From Eqs (\ref{eq:inj_rate}) and (\ref{eq:nu_diffuse_flux}), one obtains 
\begin{eqnarray}
\varepsilon_{\nu_i}^2 \Phi_{\nu_i} \sim &0.59& \times 10^{-8} \ {\rm GeV \ cm^{-2} \ s^{-1} \ sr^{-1}} (\xi_z/3)(18/{\cal C}) \notag \\
&\times& f_{pp} \xi_{cr, -1} \overline{E}_{gms, 58.5} {\cal R}_{gms, -4}
\end{eqnarray}
which can be $\sim 20\mbox{-}60 \ \%$ of the flux observed by IceCube for $\xi_{cr} = 0.1$ 
(See Fig. \ref{fig2}). 
Note that the observed local value, $R_{gms}$ still has a factor of a few uncertainty, 
and the merger  rate can evolve with redshift as $\propto (1+z)^\alpha$ with $0 \lesssim 
\alpha \lesssim 3$~\citep[e.g.,][]{Lotz_et_al_2011}. 
Such uncertainties in galaxy evolution can be absorbed in the $1 \lesssim \xi_z \lesssim 3$ factor.  
A possible PeV cutoff is consistent with the conservative estimate of the maximum CR energy, 
but the maximum neutrino energy can be larger if the UHECR acceleration is typically the case at GMSs.  
Our arguments regarding the CR acceleration and the $pp$ energy loss can be also applied to GMSs in minor mergers. 
Although the energy dissipation in a minor merger is a factor of a few smaller, the event rate is a factor of a few larger than that of a major merger. 
Given this (or a higher CR acceleration efficiency, $\xi_{cr} > 0.1$), the observed INB could be totally attributed to cosmological GMSs. 

The IGB flux can again be tightly connected to the INB flux as 
$\varepsilon_{\gamma}^2 \Phi_\gamma \approx 2 \times \varepsilon_{\nu}^2 \Phi_\nu |_{\varepsilon_\nu = 0.5 \varepsilon_\gamma}$.
Following the same reasoning as in \cite{Murase_et_al_2013}, we see here that a fraction $\gtrsim 10 \ \%$ of the observed IGB at $\sim 100 \ \rm GeV$ can be attributed to GMSs, 
for a slope of $\Gamma \approx 2.0$. 

\section{Discussion} 
As we showed above, given the concordance cosmology and a CR acceleration efficiency of $10 \ \%$, 
GMSs can provide at least $\sim 20\mbox{-}60 \ \%$ of the observed INB. 
This scenario can be indirectly tested by detecting gamma rays from the local merging galaxies using CTA. 
Within the CTA detection horizon, $d_L \lesssim 100 \ \rm Mpc$, one can expect $\approx {\cal R}_{gms} \times t_{dyn} \times (4\pi d_L{}^3/3) \lesssim 10$ such systems. 

Two of the interesting candidates are the ``Taffy" colliding galaxy pairs UGC 12914/5 
and UGC 813/6~\citep{Condon_et_al_1993, Candon_et_al_2002}, located at $d_{L} = 60 
\ \rm Mpc$ and $69 \ \rm Mpc$, respectively.  In both cases, gas components of $\sim 10^{9\mbox{-}10} 
\ M_\odot$ collided with a velocity of $\sim 600 \ \rm km \ s^{-1}$ a few $10 \ \rm Myr$ ago. 
The bridge region between the galaxy pairs filled with shocked gases, and shows strong synchrotron 
emission which may come from non-thermal electrons accelerated by the GMS~\citep{Lisenfeld_Volk_2010}.   
Another interesting candidate is VV 114~\citep{VV_2001}, which is a gas-rich merging galaxy pair 
with a core separation of $6 \ \rm kpc$, located at $d_L \sim 77 \ \rm Mpc$ from the Earth~
\citep{Soifer_et_al_1987}.  The galaxy interaction may have already triggered the starburst and 
active galactic nucleus (AGN) activity in this system~\citep{Iono_et_al_2004}.

If the acceleration in GMSs is efficient, we can again expect $\sim 10$ local sources within the GZK radius $\lesssim 100 \ \rm Mpc$, which might form hotspots in the UHECR sky~\footnote{Note that CR acceleration all the way up to the GZK cutoff energy $\sim 10^{20} \ \rm eV$ is not likely in our scenario. Here we use the GZK radius as a benchmark distance.}.
The number density of GMSs is smaller than the estimated number of sources required 
inside the GZK radius $> 10^{2\mbox{-}3}$~\citep{Auger_2013}, but the flux from the merging 
galaxies could be $\sim 10 \ \%$ of the observed one. 
Such UHECRs from local GMSs can arrive at the Earth on overlapping time windows with neutrinos and gamma rays 
for an intergalactic field of $<  \rm nG$~\citep[e.g.,][]{Bhattacharje_2000}. 

Finally, we should note that the GMS scenario is competitive with, 
but not easy to separate from, other scenarios.  
A fraction of starburst and AGN activities are triggered by galaxy mergers, as observed in VV 114.  
These latter activities can also inject a sufficient amount of CRs~\citep[e.g.,][]{Tamborra_et_al_2014}.
An identification in terms of morphology is possible based on extensive computational simulations of merger histories.
A more promising way to discriminate between the contributions from GMSs and those of 
starbursts may be to select for samples with relatively low far-infrared/radio flux ratios, 
which is observed in Taffy galaxies~\citep{Drzazga_et_al_2011}. 


\acknowledgments
The authors thank the anonymous referee for valuable comments and Xuewen Liu, Kohta Murase, and Hidenobu Yajima for discussions. 
This work is supported by NASA NNX13AH50G. 


\end{document}